\documentclass[preprint,aps,prd,showpacs,preprintnumbers]{revtex4-1}

\usepackage{graphicx}
\usepackage{amsmath,amsfonts,amssymb}
\usepackage{latexsym}
\usepackage[utf8]{inputenc}
\usepackage{xcolor}
\usepackage{color}
\begin{document}

\title{From quantum unstable systems to the decaying dark energy: Cosmological implications}
\author{Aleksander Stachowski}
\email{aleksander.stachowski@doctoral.uj.edu.pl}
\affiliation{Astronomical Observatory, Jagiellonian University, Orla 171, 30-244 Krakow, Poland}
\author{Marek Szyd{\l}owski}
\email{marek.szydlowski@uj.edu.pl}
\affiliation{Astronomical Observatory, Jagiellonian University, Orla 171, 30-244 Krakow, Poland}
\affiliation{Mark Kac Complex Systems Research Centre, Jagiellonian University, {\L}ojasiewicza 11, 30-348 Krak{\'o}w, Poland}
\author{Krzysztof Urbanowski}
\email{K.Urbanowski@if.uz.zgora.pl}
\affiliation{Institute of Physics, University of Zielona G{\'o}ra, Prof. Z. Szafrana 4a, 65-516 Zielona G{\'o}ra, Poland}

\begin{abstract}
We consider a cosmology with decaying metastable dark energy and assume that a decay process of this metastable dark energy is a quantum decay process. Such an assumption implies among others that the evolution of the Universe is irreversible and violates the time reversal symmetry. We show that if to replace the cosmological time $t$ appearing in the equation describing the evolution of the Universe by the Hubble cosmological scale time, then we obtain time dependent $\Lambda (t)$ in the form of the series of even powers of the Hubble parameter $H$: $\Lambda (t) = \Lambda (H)$. Out special attention is focused on radioactive like exponential form of the decay process of the dark energy and on the consequences of this type decay.
\end{abstract}

\maketitle

\section{Introduction}

In the explanation of the Universe, we encounter the old problem of the cosmological constant, which is related with understanding: why is the vacuum energy so small? Because of a cosmological origin of the cosmological constant one must also address another problem. Namely, it is connected with our understanding, not only with a question why the vacuum energy is not only small, but also, as current Type Ia supernova observations to indicate, why the present mass density of the universe has the same order of magnitude? \cite{Weinberg:2000yb}.

Both mentioned cosmological constant problems can be considered in the framework of the extension of the standard cosmological $\Lambda$CDM model in which the cosmological constant (naturally interpreted as the vacuum energy) is running and its value is changing during the cosmic evolution.

Results of many recent observations leads to the conclusion that our Universe is in an accelerated expansion phase \cite{Ade:2015rim}. This acceleration can be explained as a result of a presence of dark energy. A detailed analysis of results of recent observations shows that there is a tension between local and primordial measurements of cosmological parameters \cite{Ade:2015rim}. It appears that this tension may be connected with dark energy evolving in time \cite{DiValentino:2017rcr}. This paper is a contribution to the discussion of the nature of the dark energy. We consider the hypothesis that dark energy depends on time, $\rho_{\text{de}} = \rho_{\text{de}}(t)$ and it is metastable: We assume that it decays with the increasing time $t$ to $\rho_{\text{bare}}$: $\rho_{\text{de}}(t) \to \rho_{\text{bare}} \neq 0$ as $t \to \infty$. The idea that vacuum energy decays was considered in many papers (see e.g. \cite{Krauss:2007rx,Krauss:2008pt}). Shafieloo et al. \cite{Shafieloo:2016bpk} assumed that $\rho_{\text{de}}(t)$ decays according to the radioactive exponential decay law. Unfortunately, such an assumption is not able to reflect all the subtleties of evolution in the time of the dark energy and its decay process. It is because the creation of the Universe is a quantum process. Hence the metastable dark energy can be considered as the value of the scalar field at the false vacuum state and therefore the decay of the dark energy should be considered as a quantum decay process. The radioactive exponential decay law does not reflect correctly all phases of the quantum decay process. In general, analysing quantum decay processes one can distinguish the following phases \cite{Fonda:1978dk,Peshkin:2017elo}: (i) the early time initial phase, (ii) the canonical or exponential phase (when the decay law has the exponential form), and (iii) the late time non-exponential phase. The first phase and the third one are missed when one considers the radioactive decay law only. Simply they are invisible to the radioactive exponential decay law. For example, the theoretical analysis of quantum decay processes shows that at late times the survival probability of the system considered in its initial state (i.e. the decay law) should tends to zero as $t \to \infty$ much more slowly than any exponential function of time and that as a function of time it has the inverse power--like form at this regime of time \cite{Fonda:1978dk,Khalfin:1957cdt}. So, all implications of the assumption that the decay process of the dark energy is a quantum decay process can be found only if to apply a quantum decay law to describe decaying metastable dark energy. This idea was used in \cite{Szydlowski:2017wlv}, where the assumption made in \cite{Shafieloo:2016bpk} that $\rho_{\text{de}}(t)$ decays according to the radioactive exponential decay law was improved by replacing that radioactive decay law by the survival probability ${\cal P}(t)$, that is by the decay law derived assuming that the decay process is a quantum process.

This is the place where one has to emphasize that the use of the assumption that dark energy depends on time and is decaying during time evolution leads to the conclusion that such a process is irreversible and violates a time reversal symmetry. (Consequences of this effect will be analysed in next sections of this paper). Note that the picture of the evolving Universe, which results from the solutions of the Einstein equations completed with quantum corrections appearing as the effect of treating the false vacuum decay as a quantum decay process, is consistent with the observational data. The evolution starts from the early time epoch with the running $\Lambda (t)$ and then it goes to the final accelerating phase expansion of the Universe. In such a scenario the standard cosmological $\Lambda$CDM model emerges from the quantum false vacuum state of the Universe.

The paper is organised as follows: In Section~2 one finds a short introduction into a formalism necessary for considering decaying dark energy as a quantum decay process.
Cosmological implications of a decaying dark energy are considered in Section~3.
Section~4 contains conclusions.

\section{Decay of a dark energy as a quantum decay process}

In the quantum decay theory of unstable systems, properties of the survival amplitudes
\begin{equation}
{\cal A}(t) = \langle \phi| \phi(t)\rangle
\label{amp}
\end{equation}
are usually analysed. Here a vector $|\phi\rangle$ represent the unstable state of the system considered and $| \phi (t)\rangle $ is the solution of the Schr\"{o}dinger
equation
\begin{equation}
i \hbar \frac{\partial}{\partial t} |\phi(t)\rangle = \mathfrak{H}
|\phi (t)\rangle. \label{Sch}
\end{equation}
The initial condition for Eq.~(\ref{Sch}) in the case considered is usually assumed to be
\begin{equation}
| \phi (t = t_{0} \equiv 0) \rangle \stackrel{\rm def}{=}
| \phi\rangle, \label{init0}
\end{equation}
or equivalently
\begin{equation}
{\cal A}(0) = 1. \label{A(0)}
\end{equation}
In Eq.~(\ref{Sch}) $\mathfrak{H}$ denotes the complete (full), self-adjoint Hamiltonian of the system. We have $| \phi (t) \rangle = \exp\,[-\frac{i}{\hbar}t\mathfrak{H}] | \phi\rangle$. It is not difficult to see that this property and hermiticity of $H$ imply that
\begin{equation}
({\cal A}(t))^{\ast} ={\cal A}(-t). \label{amp-ast}
\end{equation}
Therefore, the decay probability of an unstable state (usually called the decay law), i.e., the probability for a quantum system to remain at time $t$ in its initial state $| \phi (0)\rangle \equiv |\phi\rangle$
\begin{equation}
{\cal P} (t) \stackrel{\rm def}{=} |{\cal A}(t)|^{2} \equiv
{\cal A}(t)\,({\cal A}(t))^{\ast}, \label{P(t)}
\end{equation}
must be an even function of time \cite{Fonda:1978dk}
\begin{equation}
{\cal P}(t) = {\cal P} ( - t). \label{even}
\end{equation}

This last property suggests that in the case of the unstable states prepared at some instant $t_{0}$, say $t_{0} = 0$, the initial condition (\ref{init0}) for the evolution equation (\ref{Sch}) should be formulated more precisely. Namely, from (\ref{even}) it follows that the probabilities of finding the system in the decaying state $|\phi\rangle$ at the instant, say $t=T \gg t_{0} \equiv 0$, and at the instant $t =-T$ are the same. Of course, this can never occur. In almost all experiments in which the decay law of a given unstable subsystem system is investigated this particle is created at some instant of time, say $t_{0}$, and this instant of time is usually considered as the initial instant for the problem. From the property (\ref{even}) it follows that the instantaneous creation of the unstable subsystem system (e.g. a particle, or an excited quantum level and so on) is practically impossible. For the observer, the creation of this object (i.e., the preparation of the state, $|\phi\rangle$, representing the decaying subsystem system) is practically instantaneous. What is more, using suitable detectors he is usually able to prove that it did not exist at times $t < t_{0}$. Therefore, if one looks for the solutions of the Schr\"{o}dinger equation (\ref{Sch}) describing properties of the unstable states prepared at some initial instant $t_{0}$ in the system, and if one requires these solutions to reflect situations described above, one should complete initial conditions (\ref{init0}), (\ref{A(0)}) for Eq.~(\ref{Sch}) by assuming additionally that
\begin{equation}
|\phi (t < t_{0})\rangle = 0 \quad \text{or} \quad
{\cal A}(t)(t < t_{0}) = 0. \label{init01}
\end{equation}
Equivalently within the problem considered, one can use initial conditions (\ref{init0}), (\ref{A(0)}) and to assume that time $t$ may vary from $t = t_{0} > - \infty$ to $t = + \infty$ only, that is that $t \in \mathbb{R}^{+}$.

Note that canonical (that is a classical radioactive) decay law ${\cal P}_{c}(t) = \exp\,[-\frac{t}{\tau_{0}}]$, (where $\tau_{0}$ is a lifetime), does not satisfy
the property (\ref{even}), which is valid only for the quantum decay law ${\cal P}(t)$. What is more, from (\ref{amp-ast}) and (\ref{P(t)}) it follows that at very early times, i. e. at the Zeno times (see \cite{Fonda:1978dk,Urbanowski:1994epq})
\begin{equation}
{\frac{\partial {\cal P}(t)}{\partial t} \vline}_{t=0} = 0, \label{dP-0}
\end{equation}
which implies that
\begin{equation}
{\cal P}(t) > e^{\textstyle{-\frac{t}{\tau_{0}}}} \stackrel{\rm def}{=} {\cal P}_{c}(t)\quad \text{for}\quad t \to 0. \label{P>Pc}
\end{equation}
So at the Zeno time region the quantum decay process is much slower than any decay process described by the canonical (or classical) decay law ${\cal P}_{c}(t)$.

Now let us focus an attention on the survival amplitude ${\cal A}(t)$. An unstable state $|\phi\rangle$ can be modeled as a wave packets using solutions of the following eigenvalue equation $\mathfrak{H}|E\rangle = E|E\rangle$, where $ E\in \sigma_{c}(\mathfrak{H})$, and $\sigma_{c}(\mathfrak{H})$ denotes a continuum spectrum of $\mathfrak{H}$. Eigenvectors $|E\rangle$ are normalised as usual: $\langle E|E'\rangle = \delta(E- E')$. Using vectors $|E\rangle$ we can model an unstable state as the following wave--packet
\begin{equation}
 |\phi\rangle \equiv |\phi\rangle
= \int_{E_{\text{min}}}^{\infty}\,c(E)\, |E\rangle\,dE, \label{phi}
\end{equation}
where
expansion coefficients $c(E)$ are functions of the energy $E$ and $E_{\text{min}}$ is the lower bound of the spectrum $\sigma_{c}(\mathfrak{H})$ of $\mathfrak{H}$. The state $|\phi\rangle$ is normalised $\langle \phi|\phi\rangle = 1$, which means that it has to be $\int_{E_{\min}}^{\infty}|c(E)|^{2}\,dE = 1$. Now using the definition of the survival amplitude ${\cal A}(t)$ and the expansion (\ref{phi}) we can find ${\cal A}(t)$, which takes the following form within the formalism considered,
\begin{equation}
{\cal A}(t) \equiv {\cal A}(t - t_{0}) = \int_{E_{\text{min}}}^{\infty} \omega(E)\;
e^{-\,i\,E\,(t-t_{0})}\,d{E},
\label{a-spec}
\end{equation}
where $\omega(E) \equiv |c(E)|^{2} > 0$ and $\omega (E)\,dE$ is the probability to find the energy of the system in the state $|\phi\rangle$ between $E$ and $E\,+\,dE$. The last relation (\ref{a-spec}) means that the survival amplitude ${\cal A}(t)$ is a Fourier transform of an absolute integrable function $\omega (E)$. If to apply the Riemann-Lebesgue Lemma to the integral (\ref{a-spec}) then one concludes that there must be ${\cal A}(t) \to 0$ as $t \to \infty$. This property and the relation (\ref{a-spec}) are an essence of the Fock--Krylov theory of unstable states \cite{Krylov:1947tmi,Fock:1978fqm}.

As it is seen from (\ref{a-spec}), the amplitude ${\cal A}(t)$, and thus the decay law ${\cal P}(t)$ of the unstable state $|\phi\rangle$, are completely determined by the density of the energy distribution $\omega(E)$ for the system in this state \cite{Krylov:1947tmi,Fock:1978fqm} (see also: \cite{Khalfin:1957cdt,Fonda:1978dk,Kelkar:2010qn,Martorell:2009qpd,Torrontegui:2010qdl,Garcia-Calderon:2008spm,Giraldi:2015ldu,Giraldi:2016zom}.

In the general case the density $\omega(E)$ possesses properties analogous to the scattering amplitude, i.e., it can be decomposed into a threshold factor, a pole-function $P(E)$ with a simple pole and a smooth form factor $F(E)$. There is $\omega(E)= {\it\Theta}(E-E_{\text{min}})\,(E-E_{\text{min}})^{\alpha_{l}}\,P(E)\,F(E) $, where $\alpha_{l}$ depends on the angular momentum $l$ through $\alpha_{l} = \alpha + l$, \cite{Fonda:1978dk} (see equation (6.1) in \cite{Fonda:1978dk}), $0 \leq \alpha <1$)and ${\it\Theta}(E)$ is a step function: ${\it\Theta}(E) = 0$ for $E \leq 0$ and ${\it\Theta}(E) = 1$ for $E>0$. The simplest choice is to take $\alpha = 0$, $l=0$, $F(E) = 1$ and to assume that $P(E)$ has a Breit--Wigner (BW) form of the energy distribution density. (The mentioned Breit--Wigner distribution was found when the cross--section of slow neutrons was analysed \cite{Breit:1936zzb}). It turns out that the decay curves obtained in this simplest case are very similar in form to the curves calculated for the above described more general $\omega (E)$, (see \cite{Kelkar:2010qn} and analysis in \cite{Fonda:1978dk}). So to find the most typical properties of the decay process it is sufficient to make the relevant calculations for $\omega (E)$ modelled by the the Breit--Wigner distribution of the energy density
$
\omega (E) \equiv \omega_{\text{BW}}(E) \stackrel{\text{def}}{=} \frac{N}{2\pi}\, {\it\Theta} (E - E_{\text{min}}) \
\frac{{\it\Gamma}_{0}}{(E-E_{0})^{2} +
(\frac{{\it\Gamma}_{0}}{2})^{2}},
$
where $N$ is a normalization constant.
The parameters $E_{0}$ and ${\it\Gamma}_{0}$ correspond to the energy of the system in the unstable state and its decay rate at the exponential (or canonical) regime of the decay process. $E_{\text{min}}$ is the minimal (the lowest) energy of the system. Inserting $\omega_{\text{BW}}(E)$ into formula (\ref{a-spec}) for the amplitude ${\cal A}(t)$ and assuming for simplicity that $t_{0} = 0$, after some algebra one finds that
\begin{equation}
{\cal A}(t) = \frac{N}{2\pi}\,
e^{\textstyle{ - \frac{i}{\hbar} E_{0}t }}\, I_{\beta}\left(\frac{{\it\Gamma}_{0} t}{\hbar}\right), \label{I(t)a}
\end{equation}
where
\begin{equation}
I_{\beta}(\tau) \stackrel{\rm def}{=}\int_{-\beta}^{\infty}
 \frac{1}{\eta^{2}
+ \frac{1}{4}}\, e^{\textstyle{ -i\eta\tau}}\,d\eta. \label{I(t)}
\end{equation}
Here $\tau = \frac{{\it\Gamma}_{0}\,t}{\hbar} \equiv \frac{t}{\tau_{0}}$, $\tau_{0}$ is the lifetime, $\tau_{0} = \frac{\hbar}{{\it\Gamma}_{0}}$, and $\beta = \frac{E_{0} - E_{min}}{{\it\Gamma}_{0}}\,>\,0$. The integral $I_{\beta}(\tau)$ has the following structure
\begin{equation}
I_{\beta}(\tau) = I_{\beta}^{\text{pole}}(\tau) + I_{\beta}^{L}(\tau), \label{Ip+IL}
\end{equation}
where
\begin{equation}
 I_{\beta}^{\text{pole}}(\tau) = \int_{- \infty}^{\infty}
 \frac{1}{\eta^{2}
+ \frac{1}{4}}\,\; e^{\textstyle{ -
i \eta \tau}} \; d\eta \equiv 2 \pi\,e^{\textstyle{-\,\frac{\tau}{2}}}, \label{Ip}
 \end{equation}
 and
 \begin{equation}
 I_{\beta}^{L}(\tau) = -
 \int_{+ \beta}^{\infty}
 \frac{1}{\eta^{2}
+ \frac{1}{4}}\,\; e^{\textstyle{ +
i \eta \tau}} \; d\eta. \label{IL}
 \end{equation}
(The integral $I_{\beta}^{L}(\tau)$ can be expressed in terms of the integral--exponential function \cite{Sluis:1991dqs,Urbanowski:2006mw,Urbanowski:2009lpe,Raczynska:2018ofh} (for a definition, see \cite{Olver:2010hmf,Abramowitz:1964hmf})). The result (\ref{Ip+IL}) means that there is a natural decomposition of the survival amplitude ${\cal A}(t)$ into two parts
\begin{equation}
{\cal A}(t) = {\cal A}_{c}(t) + {\cal A}_{L}(t), \label{Ac+AL}
\end{equation}
where
\begin{equation}
{\cal A}_{c}(t) = \frac{N}{2\pi}\,
e^{\textstyle{ - \frac{i}{\hbar} E_{0}t }}\,I_{\beta}^{\text{pole}}\left(\frac{{\it\Gamma}_{0} t}{\hbar}\right) \equiv N\,e^{\textstyle{ - \frac{i}{\hbar} E_{0}t }}\,
e^{\textstyle{- \frac{{\it\Gamma}_{0}\,t}{2}}}, \label{Ac(t)}
\end{equation}
and
\begin{equation}
{\cal A}_{L}(t) = \frac{N}{2\pi}\,
e^{\textstyle{ - \frac{i}{\hbar} E_{0}t }}\,I_{\beta}^{L}\left(\frac{{\it\Gamma}_{0} t}{\hbar}\right), \label{AL(t)}
\end{equation}
and ${\cal A}_{c}(t)$ is the canonical part of the amplitude ${\cal A}(t)$ describing the pole contribution into ${\cal A}(t)$ and ${\cal A}_{L}(t)$ represents the remaining part of ${\cal A}(t)$.

From the decomposition (\ref{Ac+AL}) it follows that in the general case within the model considered the survival probability (\ref{P(t)}) contains the following parts
\begin{eqnarray}
{\cal P}(t) = |{\cal A}(t)|^{2} &\equiv& |{\cal A}_{c}(t) + {\cal A}_{L}(t)|^{2} \nonumber \\
&=& |{\cal A}_{c}(t)|^{2} \,+ \,2\,\Re\,[{\cal A}_{c}(t)\,( {\cal A}_{L}(t) )^{\ast}]\,+\,|{\cal A}_{L}(t)|^{2}. \label{Ac+AL-2}
\end{eqnarray}
This last relation is especially useful when one looks for a contribution of a late time properties of the quantum unstable system into the survival amplitude.

The late time form of the integral $I_{\beta}^{L}(\tau)$ and thus the late time form of the amplitude ${\cal A}_{L}(t)$ can be relatively easy to find using analytical expression for ${\cal A}_{L}(t)$ in terms of the integral--exponential functions or simply performing the integration by parts in (\ref{IL}). One finds for $t \to \infty$ (or $\tau \to \infty$) that the leading term of the late time asymptotic expansion of the integral $I_{\beta}^{L}(\tau)$ has the following form
\begin{eqnarray}
I_{\beta}^{L}(\tau) &\simeq & -\, \frac{i}{\tau}\,\frac{e^{\textstyle{i\beta \tau}}}{\beta^{2} + \frac{1}{4}}\,
 + \ldots , \;\;\;(\tau \to \infty). \label{IL-as}
\end{eqnarray}
Thus inserting (\ref{IL-as}) into (\ref{AL(t)}) one can find late time form of ${\cal A}_{L}(t)$.

As it was mentioned we consider the hypothesis that a dark energy depends on time, $\rho_{\text{de}} = \rho_{\text{de}}(t)$ and decays with the increasing time $t$ to $\rho_{\text{bare}}$: $\rho_{\text{de}}(t) \to \rho_{\text{bare}} \neq 0$ as $t \to \infty$. We assume that it is a quantum decay process. The consequence of this assumption is that we should consider $\rho_{\text{de}}(t_{0})$ (where $t_{0}$ is the initial instant) as the energy of an excited quantum level (e.g. corresponding to the false vacuum state) and the energy density $\rho_{\text{bare}}$ as the energy corresponding to the true lowest energy state (the true vacuum) of the system considered. Our hypothesis means that $(\rho_{\text{de}}(t) - \rho_{\text{bare}}) \to 0$ as $t \to \infty$. As it was said we assumed that that the decay process of the dark energy is a quantum decay process: From the point of view of the quantum theory of decay processes this means that $\lim_{t \to \infty} (\rho_{\text{de}}(t) - \rho_{\text{bare}}) = 0$ according to the quantum mechanical decay law. Therefore if to define
\begin{equation}
\tilde{\rho}_{\text{de}}(t) \stackrel{\rm def}{=} \rho_{\text{de}}(t) - \rho_{\text{bare}},
\end{equation}
our assumption means that the decay law for $\tilde{\rho}_{\text{de}}(t)$ has the following form (see \cite{Szydlowski:2017wlv}
\begin{eqnarray}
\tilde{\rho}_{\text{de}}(t) &=& \tilde{\rho}_{\text{de}}(t_{0})\,{\cal P}(t) \nonumber \\ &\equiv& \tilde{\rho}_{\text{de}}(t_{0})\left(|{\cal A}_{c}(t)|^{2} \,+ \,2\,\Re\,[{\cal A}_{c}(t)\,( {\cal A}_{L}(t) )^{\ast}]\,+\,|{\cal A}_{L}(t)|^{2}\right), \label{rho-tilde(t)}
\end{eqnarray}
where ${\cal P}(t)$ is given by the relation (\ref{P(t)}), or equivalently, our assumption means that the decay law for $\tilde{\rho}_{\text{de}}(t)$ has the following form (compare \cite{Szydlowski:2017wlv})
\begin{equation}
\rho_{\text{de}}(t) \equiv \rho_{\text{bare}} + \tilde{\rho}_{\text{de}}(t_{0})\left(|{\cal A}_{c}(t)|^{2} \,+ \,2\,\Re\,[{\cal A}_{c}(t)\,( {\cal A}_{L}(t) )^{\ast}]\,+\,|{\cal A}_{L}(t)|^{2}\right), \label{rho(t)}
\end{equation}
where $\tilde{\rho}_{\text{de}}(t_{0}) = (\rho_{\text{de}}(t_{0}) - \rho_{\text{bare}})$ and ${\cal P}(t)$ is replaced by (\ref{Ac+AL-2}). Taking into account the standard relation between $\rho_{\text{de}}$ and the cosmological constant $\Lambda$ we can write that
\begin{equation}
\Lambda_{\text{eff}}(t) \equiv \Lambda_{\text{bare}} + \tilde{\Lambda}(t_{0})\left(|{\cal A}_{c}(t)|^{2} \,+ \,2\,\Re\,[{\cal A}_{c}(t)\,( {\cal A}_{L}(t) )^{\ast}]\,+\,|{\cal A}_{L}(t)|^{2}\right), \label{Lambda(t)}
\end{equation}
where $\tilde{\Lambda}(t_{0}) \equiv \tilde{ \Lambda}_{0} = (\Lambda(t_{0}) - \Lambda_{\text{bare}})$. Thus within the considered case using the definition (\ref{P(t)}), or the relation  (\ref{Ac+AL-2}) we can determine changes in time of the dark energy density $\rho_{\text{de}}(t)$ (or running $\Lambda (t)$) knowing the general properties of survival amplitude ${\cal A}(t)$.

The above described approach is self consistent if to identify $\rho_{\text{de}}(t_{0})$ with the energy $E_{0}$ of the unstable system divided by the volume $V_{0}$ (where $V_{0}$ is the volume of the system at $t=t_{0}$): $\rho_{\text{de}}(t_{0}) \equiv \rho_{\text{de}}^{\text{qft}} \stackrel{\rm def}{=} \rho_{\text{de}}^{0} = \frac{E_{0}}{V_{0}}$ and $ \rho_{\text{bare}} =\frac{E_{\text{min}}}{V_{0}}$. Here $\rho_{\text{de}}^{\text{qft}}$ is the vacuum energy density calculated using quantum field theory methods. In such a case
\begin{equation}
\beta = \frac{E_{0} - E_{\text{min}}}{{\it\Gamma}_{0}} \equiv \frac{\rho_{\text{de}}^{0} - \rho_{\text{bare}}}{{\gamma}_{0}} > 0, \label{beta-rho}
\end{equation}
(where $\gamma_{0} = {\it\Gamma}_{0}/V_{0}$), or equivalently, $ {\it\Gamma}_{0}/V_{0} \equiv \frac{\rho_{\text{de}}^{0} - \rho_{\text{bare}}}{\beta}$.

\section{Cosmological implications of decaying vacuum}

Let us consider cosmological implications of the parameter $\Lambda$ with the time parameterized decaying part, derived in the previous section, in the form
\begin{equation}
\Lambda \equiv \Lambda_{\text{eff}}(t) = \Lambda_\text{bare}+\delta\Lambda(t),
\end{equation}
where $\delta \Lambda(t)$ describes quantum corrections and it is given by a series with respect to $\frac{1}{t}$, i.e.
\begin{equation}
\delta\Lambda(t)=\sum^\infty_{n=1}\alpha_{2n}\left(\frac{1}{t}\right)^{2n},\label{sum}
\end{equation}
where $t$ is the cosmological scale time and the functions $\Lambda_{\text{eff}}(t)$ and $\delta \Lambda(t)$ have a reflection symmetry with respect to the cosmological time $\delta \Lambda(-t)=\delta \Lambda(t)$. The next step in deriving dynamical equations for the evolution of the Universe is to consider this parameter as a source of gravity which contributes to the effective energy density, i.e.
\begin{equation}
3H(t)^2=\rho_\text{m}(t)+\rho_\text{de}(t),\label{friedmann}
\end{equation}
where $\rho_\text{de}(t)$ is identify as the energy density of the quantum decay process of vacuum
\begin{equation}
\rho_\text{de}(t)=\Lambda_\text{bare}+ \delta\Lambda(t).
\end{equation}
The Einstein field equation for the FRW metric reduces to
\begin{equation}
\frac{dH(t)}{dt}=-\frac{1}{2}(\rho_\text{eff}(t)+p_\text{eff}(t))=-\frac{1}{2}(\rho_\text{m}(t)+0+\rho_\text{de}(t)-\rho_\text{de}(t)),
\end{equation}
where $\rho_\text{eff}=\rho_\text{m}+\rho_\text{de}$, $p_\text{eff}=0+p_\text{de}$ or
\begin{equation}
\frac{dH(t)}{dt}=-\frac{1}{2} \rho_\text{m}(t) = -\frac{1}{2}\left(3H(t)^2-\Lambda_\text{bare}-\delta \Lambda(t)\right).\label{hubble}
\end{equation}

Szydlowski et al. \cite{Szydlowski:2017wlv} considered the radioactive-like decay of metastable dark energy. For the late time, this decay process has three consecutive phases: the phase of radioactive decay, the phase of damping oscillations, and finally the phase of power law decaying. When $\beta >0$ for $t > \frac{\hbar}{\it{\Gamma}_{0}}\,\frac{2\beta}{\beta^{2} + \frac{1}{4}}$, dark energy can be described in the following form (see (\ref{rho(t)}) and \cite{Szydlowski:2017wlv})
\begin{equation}
\rho_\text{de}(t)\approx\rho_\text{bare}+\\
\epsilon \left(4\pi^2 e^{-\frac{\it{\Gamma}_0}{\hbar}t}+\frac{4\pi e^{-\frac{\it{\Gamma}_0}{2\hbar}t}\sin\left(\beta\,\frac{\it{\Gamma}_0}{\hbar}t\right)}{\left(\frac{1}{4}+{\beta^2}\right)\frac{\it{\Gamma}_0}{\hbar}t}+
\frac{1}{\left(\left(\frac{1}{4}+{\beta^2}\right)\frac{\it{\Gamma}_0}{\hbar}t\right)^2}\right),\label{darkenergy3}
\end{equation}
where $\epsilon$, $\Gamma_0$ and $\beta$ are model parameters. Equation~(\ref{darkenergy3}) results directly from (\ref{rho(t)}): One only needs to insert (\ref{IL-as}) into formula for ${\cal A}_{L}(t)$ and the result (\ref{Ac(t)}) instead of ${\cal A}_{c}(t)$ into (\ref{rho(t)}).
In this paper, we consider the first phase of decay process, in other words, the phase of radioactive (exponential) decay.

The model with the radioactive (exponential) decay of dark energy was investigated by Shafieloo et al. \cite{Shafieloo:2016bpk}. During the phase of the exponential decay of the vacuum
\begin{equation}
\frac{d\delta\Lambda(t)}{dt}=A\delta\Lambda(t),\label{lambda}
\end{equation}
where $A=\text{const}<0$ ($\delta\Lambda(t)$ is decaying).

The set of equations~(\ref{hubble}) and (\ref{lambda}) constitute a two-dimensional closed autonomous dynamical system in the form
\begin{equation}
\begin{split}
\frac{dH(t)}{dt} & =-\frac{1}{2}\left(3H(t)^2-\Lambda_\text{bare}- \delta\Lambda(t)\right),\\
\frac{d\delta\Lambda(t)}{dt} & =A\delta\Lambda(t). \label{dynamical}
\end{split}
\end{equation}

The system (\ref{dynamical}) has the time-dependent first integral in the form
\begin{equation}
\rho_\text{m}(t)=3H(t)^2-\Lambda_\text{bare} - \delta\Lambda(t). \label{friedmann2}
\end{equation}
At the finite domain, the system (\ref{dynamical}) possesses only one critical point representing the standard cosmological model (the running part of $\Lambda$ vanishes, i.e. $\delta\Lambda(t)=0$).

\begin{figure}
	\centering
	\includegraphics[width=0.7\linewidth]{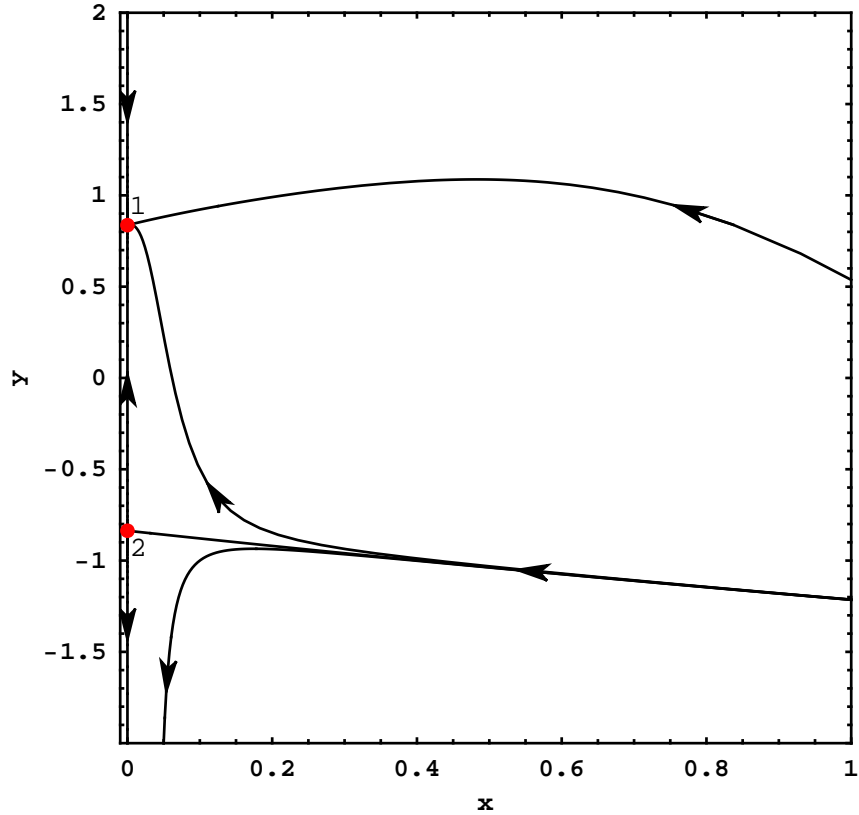}
	\caption{The phase portrait of the system (\ref{dynamical2}). Critical point 1 $\left(x=0,\ y=\frac{\sqrt{\Lambda_{\text{bare}}}}{\sqrt{3}H_0}\right)$ is the stable node and critical point 2 $\left(x=0,\ y=-\frac{\sqrt{\Lambda_{\text{bare}}}}{\sqrt{3}H_0}\right)$ is the saddle. These critical points represent the de Sitter universes. Here, $H_0$ is the present value of the Hubble function. The value of $A$ is assumed as $-1$. Note that the phase portrait is not symmetric under reflection $H \to -H$. While critical point 1 is a global attractor, only a unique separatrix reaches critical point 2.}
	\label{fig1}
\end{figure}

The system (\ref{dynamical}) can be rewritten in variables
\begin{equation}
x=\frac{\delta\Lambda(t)}{3H_0^2}, \qquad y=\frac{H(t)}{H_0}
\end{equation}
where $H_0$ is the present value of the Hubble function. Then
\begin{equation}
\begin{split}
\frac{dx}{dt} & =Ax\\
\frac{dy}{dt} & =-\frac{1}{2}\left(3y^2-3\Omega_{\Lambda_\text{bare}}-3x\right), \label{dynamical2}
\end{split}
\end{equation}
where $\Omega_{\Lambda_\text{bare}}=\frac{\Lambda_{\text{bare}}}{3H_0^2}$. The phase portrait of system (\ref{dynamical2}) is shown in Fig.~\ref{fig1}.

Szydlowski et al. \cite{Szydlowski:2017wlv} demonstrated that the contribution of the energy density of the decaying quantum vacuum possesses three disjoint phases during the cosmic evolution. The phase of exponential decay like in the radioactive decay processes is long phases in the past and future evolution. Our estimation of model parameter shows that we are living in the Universe with the radioactive decay of the quantum vacuum.

It is interesting, that during this phase, the universe violates the reflection symmetry of the time: $t\rightarrow -t$. In cosmology and generally in physics there is a fundamental problem of the origin of irreversibility in the Universe \cite{Zeh:2007pbd}. Note that in our model irreversibility is a consequence of the radioactive decay of the quantum vacuum.

In the general parameterization (\ref{sum}), of course, there is present the symmetry of changing $t\rightarrow -t$ and this symmetry is also in a one-dimensional non-autonomous dynamical system describing the evolution of the Universe
\begin{equation}
\frac{dH(t)}{dt} =-\frac{1}{2}\left(3H(t)^2-\Lambda_\text{bare}-\sum^\infty_{n=1}\alpha_{2n}t^{-2n}\right).\label{hubble2}
\end{equation}

In cosmology, especially in quantum cosmology, the analysis of the concept of time seems to be the key for the construction an adequate quantum gravity theory, which we would like to apply to the description of early Universe.

\begin{figure}
	\centering
	\includegraphics[width=0.7\linewidth]{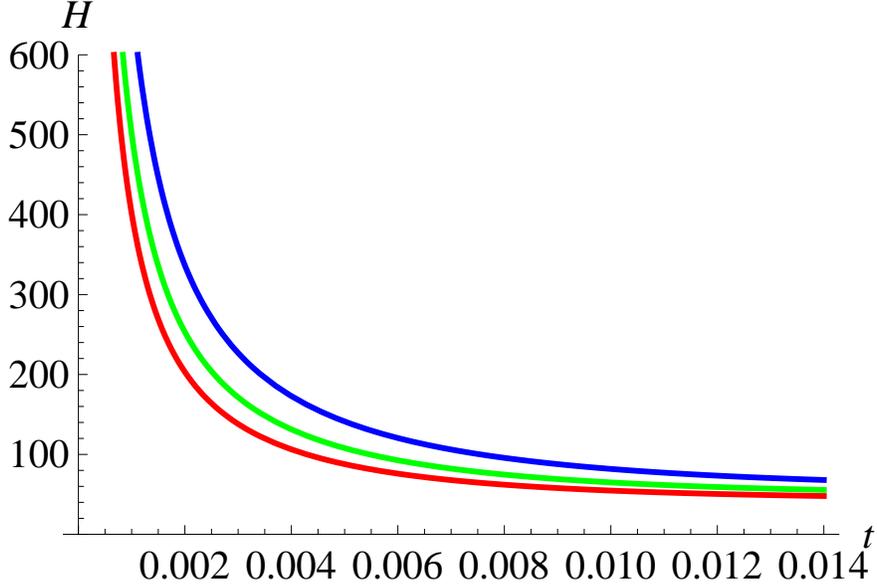}
	\caption{The diagram of the evolution of the Hubble function with respect of the cosmological time $t$, which is described by Eq.~(\ref{hubble3}) with $\alpha_{21}\neq 0$ and $\alpha_{2n}=0$ for every $n>1$. For illustration, two example values of the parameter $\alpha_{21}=$ are chosen: $-1$ and $-2$. The top blue curve describes the evolution of the Hubble function in the $\Lambda$CDM model. The middle curve describes one for $\alpha_{21}=-1$ and the bottom red curve describes one for $\alpha_{21}=-2$. The Hubble function is expressed in $\frac{\text{km}}{\text{s}\ \text{Mpc}}$ and the cosmological time $t$ is expressed in $\frac{\text{s}\ \text{Mpc}}{\text{km}}$.}
	\label{fig2}
\end{figure}

The good approximation of Eq.~(\ref{hubble2}) is to replace in it the cosmological time by the Hubble cosmological scale time
\begin{equation}
t_\text{H}=\frac{1}{H}.
\end{equation}
In the consequence, the parameterization (\ref{sum}) can be rewritten in the new form
\begin{equation}
\delta\Lambda(t)=\delta\Lambda(H(t))=\sum^\infty_{n=1}\alpha_{2n}H(t)^{2n}.
\end{equation}
After putting this form into (\ref{hubble2}), we obtain dynamical system in an autonomous form with the preserved symmetry of time $t\rightarrow -t$, $H\rightarrow -H$. In Fig.~\ref{fig2} it is presented a diagram of the evolution of the Hubble function obtained from Eq.~(\ref{hubble2}) and the following one-dimensional dynamical system
\begin{equation}
\frac{dH(t)}{dt} =-\frac{1}{2}\left(3H(t)^2-\Lambda_\text{bare}-\sum^\infty_{n=1}\alpha_{2n}H(t)^{2n}\right).\label{hubble3}
\end{equation}
For the existence of the de Sitter global attractor as $t\rightarrow\infty$ asymptotically a contribution coming from the decaying part of $\delta\Lambda(H(t))=\sum^\infty_{n=1}\alpha_{2n}H(t)^{2n}$ should be vanishing.

\begin{figure}
	\centering
	\includegraphics[width=0.7\linewidth]{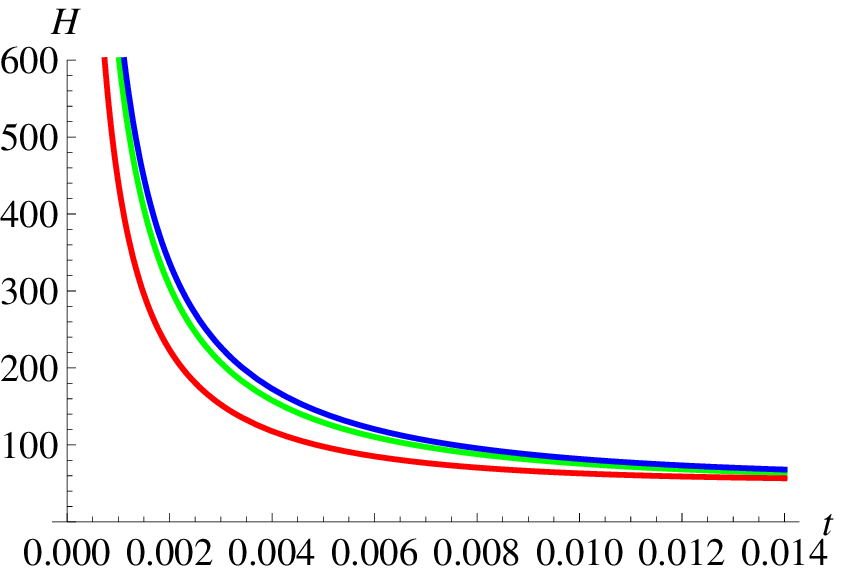}
	\caption{The diagram of the evolution of the Hubble function with respect of the cosmological time $t$, which is described by Eqs (\ref{hubble2}) and (\ref{hubble3}) with $\alpha_{21}\neq 0$ and $\alpha_{2n}=0$ for every $n>1$. For illustration, the value of the parameter $\alpha_{21}=$ is chosen as $-0.3$. The top blue curve describes the evolution of the Hubble function in the $\Lambda$CDM model. The middle curve describes one for one described Eq.~(\ref{hubble3}) and the bottom red curve describes one for Eq.~(\ref{hubble2}). The Hubble function is expressed in $\frac{\text{km}}{\text{s}\ \text{Mpc}}$ and the cosmological time $t$ is expressed in $\frac{\text{s}\ \text{Mpc}}{\text{km}}$. Note that, these models are not qualitatively different.}
	\label{fig3}
\end{figure}

This condition guarantees us a consistency of our model with astronomical observations of the accelerating phase of the universe \cite{Ade:2015rim}.

If all parameters $\alpha_{2n}$ for $n>1$ are equal zero then the Hubble parameter is described by the following formula
\begin{equation}
H(t)=\pm\sqrt{\frac{\rho_\text{m,0}a^{\alpha_{21}-3}+\Lambda_{bare}}{3-\alpha_{21}}}. \label{eq:44}
\end{equation}
From Eq.~(\ref{eq:44}) we can obtain the following formula for the expanding universe
\begin{equation}
a(t)=\left(\frac{\rho_\text{m,0}}{\Lambda_{bare}} \sinh \left(\frac{\sqrt{(3-\alpha_{21})\Lambda_{\text{bare}}}}{2}t\right)\right)^{\frac{2}{3-\alpha_{21}}}.\label{scale}
\end{equation}

\begin{figure}
	\centering
	\includegraphics[width=0.7\linewidth]{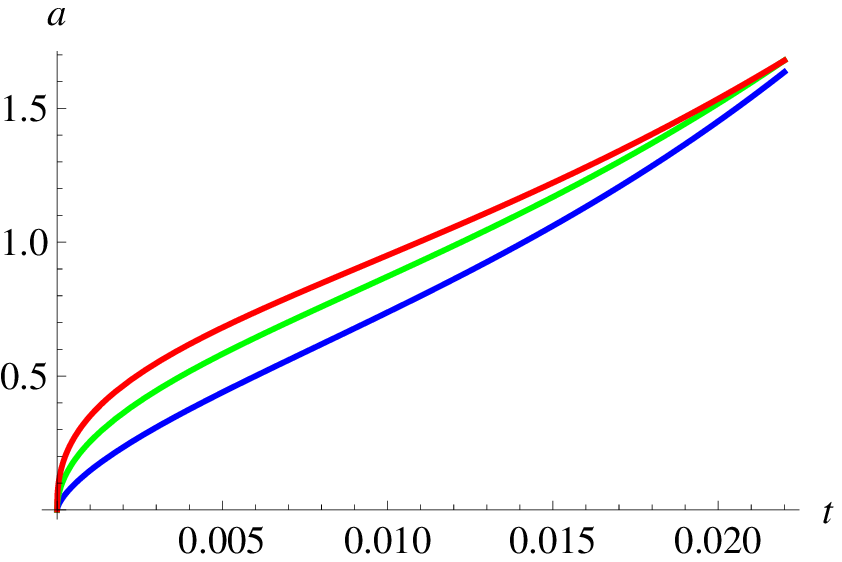}
	\caption{The diagram of the evolution of the scale factor with respect of the cosmological time $t$, which is described by Eq.~(\ref{scale}). For illustration, two example values of the parameter $\alpha_{21}=$ are chosen: $-1$ and $-2$. The bottom blue curve describes the evolution of the scale factor in the $\Lambda$CDM model. The middle curve describes one for $\alpha_{21}=-1$ and the top red curve describes one for $\alpha_{21}=-2$. The cosmological time $t$ is expressed in $\frac{\text{s}\ \text{Mpc}}{\text{km}}$.}
	\label{fig5}
\end{figure}

Fig.~\ref{fig5} presents the evolution of the scale factor, which is described by Eq.~(\ref{scale}). Eq.~(\ref{scale}) gives us the following formula
\begin{equation}
H(t)=\sqrt{\frac{\Lambda_{\text{bare}}}{3-\alpha_{21}}} \coth \left(\frac{1}{2}\sqrt{\Lambda_{\text{bare}}(3-\alpha_{21})}t\right).
\end{equation}

For comparison, the evolution of the Hubble functions derived in the $\Lambda$CDM model, the model (\ref{hubble2}) and the model (\ref{hubble3}) are presented in Fig.~\ref{fig3}.

\begin{figure}
	\centering
	\includegraphics[width=0.7\linewidth]{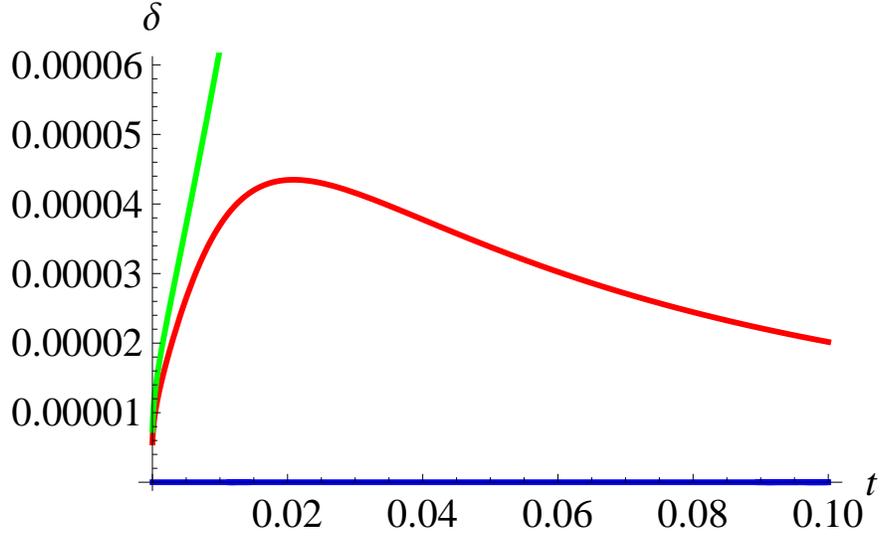}
	\caption{The diagram of the evolution of the parameter $\delta$ with respect to the cosmological time $t$. For illustration, two example values of the parameter $A$ are chosen: $A=-100\frac{\text{km}}{\text{s}\ \text{Mpc}}$ (the top red curve) and $A=-200\frac{\text{km}}{\text{s}\ \text{Mpc}}$ (the middle green curve). For comparison the $\Lambda$CDM model with the parameter $\delta = 0$ is represented by the bottom blue curve. Here, the value of $B$ parameter is equal 1. The cosmological time $t$ is expressed in $\frac{\text{s}\ \text{Mpc}}{\text{km}}$.}
	\label{fig4}
\end{figure}

In the extension of the Friedmann equation (\ref{friedmann2}) matter is contributed as well as dark energy. The total energy-momentum tensor $T^{\mu\nu}=T^{\mu\nu}_\text{m}+T^{\mu\nu}_\text{de}$ is of course conserved. However, between the matter and dark energy sectors exist an interaction---the energy density is transferred between these sectors. This process can be described by the system of equations
\begin{equation}
\begin{split}
\frac{d\rho_\text{m}(t)}{dt}+3H(t)\rho_\text{m}(t) & =-\frac{d\rho_\text{de}(t)}{dt}=-\frac{d\Lambda_{\text{eff}}(t)}{dt},\\
\frac{d\rho_\text{de}(t)}{dt} & =\frac{d\Lambda_{\text{eff}}(t)}{dt},\label{rho}
\end{split}
\end{equation}
where it is assumed that pressure of matter $p_\text{m}=0$ and $p_\text{de}=-\rho_\text{de}$. The time variability of the matter and energy density of decaying vacuum is demonstrated in Fig.~\ref{fig4}.

In the special case of radioactive decay of vacuum Eqs (\ref{rho}) reduces to
\begin{equation}
\begin{split}
\frac{d\rho_\text{m}(t)}{dt}+3H(t)\rho_\text{m}(t) & =-ABe^{Bt}=-B\delta\Lambda(t),\\
\frac{d\rho_\text{de}(t)}{dt} & =B\delta\Lambda(t)
\end{split}\label{decay}
\end{equation}
or
\begin{equation}
\begin{split}
\frac{1}{a(t)^3}\frac{d}{dt}(a(t)^3\rho_\text{m}(t)) & =-ABe^{Bt}=-B\delta\Lambda(t)\Rightarrow \rho_\text{m}(t) a(t)^3=\rho_\text{m,0}a^3_0-\int ABe^{Bt}a^3(t)dt,\\
\frac{d\rho_\text{de}(t)}{dt} & =B\delta\Lambda(t).
\end{split}
\end{equation}

Let $\rho_\text{m}(t)=\rho_{m,0}a^{-3+\delta(t)}$, where $\delta(t)$ is a deviation from the canonical scaling of dust matter. Then from Eq.~(\ref{decay}), we have
\begin{equation}
\delta(t)=\frac{\ln\frac{\rho_\text{m}(t)}{\rho_\text{m,0}}}{\ln a(t)}+3.
\end{equation}

\section{Conclusions}

From our investigation of cosmological implications of effects of the quantum decay of metastable dark energy, one can derive following results:
\begin{itemize}
\item The cosmological models with the running cosmological parameter can be included in the framework of some extension of Friedmann equation. The new ingredient in the comparison with the standard cosmological model ($\Lambda$CDM model) is that the total energy-momentum tensor is conserved and the interaction takes place between the matter and dark energy sectors. In the consequence the canonical scaling law $\rho_\text{m}\propto a^{-3}$ is modified. Because $\Lambda(t)$ is decaying ($\frac{d\Lambda}{dt}<0$) energy of matter in the comoving volume $\propto a^{3}$ is growing with time.
\item We have found that the appearance of the universal exponential contribution in energy density of the decaying vacuum can explain the irreversibility of the cosmic evolution. While the reversibility $t\rightarrow -t$ is still present in the dynamical equation describing the evolutional scenario, in the first phase of radioactive decay, this symmetry is violated.
\item We have also compared the time evolution of the Hubble function in the model under consideration (where $\Lambda(t)$ is parameterized by the cosmological time) with Sola et al. \cite{Shapiro:2009dh} parameterization by the Hubble function. Note that both parameterizations coincide if time $t$ is replaced by the Hubble scale time $t_{H}=\frac{1}{H}$. If the evolution of the Universe is invariant in the scale, i.e. the scale factor $a$ is changing in power law, then this correspondence is exact.
\end{itemize}

\end{document}